\documentclass[letterpaper]{aipproc}
\layoutstyle{6x9}

\usepackage{tabularx,amsmath,amssymb,bm}
\usepackage{citesort}

\begin{document}

\begin{flushright}
JLAB-THY-11-1164\vspace*{-0.86cm}
\end{flushright}

\title{Subleading effects in QCD global fits}


\author{Alberto Accardi}{  
  address={Hampton University, Hampton, Virginia 23668 \\
  and Jefferson Lab, Newport News, Virginia 23606} 
}

\begin{abstract}
I discuss several corrections to leading twist calculations of nucleon
structure functions which are needed to include experimental data at
large parton fractional momentum $x$ and at low scales $Q^2$ 
in global fits of parton distribution functions.
In particular I discuss the results of the CTEQ6X global fit,
and some work in progress. Topics covered include the interplay
of target mass and higher-twist corrections, the importance of nuclear
corrections for deuterium target data, and applications to the study
of quark-hadron duality. Implications for collider physics are
highlighted. 
\end{abstract}

\maketitle

%%%%%%%%%%%%%%%%%%%%%%%%%%%%%%%%%%%%%%%%%%%%
%% MAINMATTER
%%%%%%%%%%%%%%%%%%%%%%%%%%%%%%%%%%%%%%%%%%%%

Precise parton distribution functions (PDFs) at large parton fractional
momentum $x$ are vital for understanding the non perturbative
structure of the nucleon and the effects of color confinement on its
partonic constituents. For instance, the $d/u$ quark distribution ratio
near $x=1$ is very sensitive to the nature of the quark-quark forces in
the nucleon; the ratios of 
spin-polarized to spin-averaged PDFs $\Delta u/u$, and particularly
$\Delta d/d$, in the limit $x \to 1$ reflect the non perturbative
quark-gluon dynamics in the nucleon, and can shed light on the
origin of the nucleon's spin. Precise PDFs at large $x$ also have
impact in other 
areas of nuclear and high-energy physics, e.g., by allowing precise
computations of QCD background processes in searches of new physics
signals at hadron colliders, and systematic
uncertainties in neutrino oscillation experiments.

PDFs can be extracted from experimental data through global QCD fits
which combine data from many different processes
and observables, and analyze them by means of perturbative QCD
calculations
\cite{Owens-HiX2010,Guffanti-HiX2010,Radescu-HiX2010,Holt-HiX2010}. 
Currently, however, the unpolarized PDFs are well determined
only for $x \lesssim 0.5$ for valence quarks, $x \lesssim 0.3$ for
gluons, and $x\lesssim0.1$ for heavy quarks. 
To better constrain these at large $x$ it is necessary to study hard
scattering processes near kinematic thresholds, such as Deep Inelastic
Scattering (DIS) at large Bjorken invariant $x_B$ and low 4-momentum transfer
squared $Q^2$, Drell-Yan (DY) lepton pair production and electroweak
vector boson production at large rapidity. In these kinematic regimes
several corrections to leading twist perturbative QCD calculations can
become important because of the rapid fall-off of the cross section
near the kinematic boundary. Examples are target and jet mass
corrections \cite{Steffens-HiX2010}, threshold resummation
\cite{Liuti-HiX2010}, and higher-twists (HT) contributions
\cite{Glatzmaier-HiX2010,Lee-HiX2010}.   
Moreover, data taken on nuclear
targets must be corrected for nuclear effects such as shadowing,
binding, Fermi motion and nucleon off-shellness, to access the
partonic structure at the nucleon level.  
Accessing the highest values of $x$ in DIS also requires understanding
quark-hadron duality \cite{Melnitchouk-HiX2010,Malace-HiX2010} in
order to utilize data in the resonance region.

All these effects need to be incorporated in a consistent framework,
simultaneously computed for a wide range observables, and utilized in
a global PDF fit. The CTEQ6X global fit published
in Ref.~\cite{Accardi:2009br} took a first step in this program by
considering 
the combined effect of TMC and HT corrections, alongside nuclear
corrections for DIS data on deuterium targets needed for flavor
separation of the up and down quark. In this talk, I discuss the
results of this analysis and some recent work in progress which extends
it. Detailed references can be found in Ref.~\cite{Accardi:2009br}.

\section{Target mass and higher-twist corrections}

It is the usual practice in global PDF fits to place kinematic
constraints on the DIS data sets, typically $Q^2>4$ GeV$^2$ and $W^2>12$
GeV$^2$, so that only leading twist massless QCD contributions
need be considered, thereby reducing the model dependent error on the
extracted PDFs. As a byproduct of this procedure
the PDFs are directly constrained by data only in the region $x < 0.7$.
However, plentiful DIS data exist outside this region. In order to
utilize them in global fits, one needs minimally to include Target
Mass Corrections (TMCs), which scale as $M_N^2/Q^2$, with $M_N$ the nucleon
mass, and higher-twist corrections, which scale as $\lambda^2/Q^2$,
with $\lambda$ a hadronic scale describing non perturbative parton-parton
correlations inside the nucleon. 

Several methods are available in the
literature to perform TMCs, and have been reviewed by F.~Steffens
\cite{Steffens-HiX2010}. One is the well-known 
Georgi-Politzer formalism based on the
Operator Product Expansion (OPE), whose results are also reproduced in
the Covariant Parton Model. However, this formalism
suffers from the problem that it 
leads to non-zero values of the
structure function on a nucleon target in the unphysical region
$x>1$. Another prescription is a simple rescaling of the structure
function, obtained by substituting $x$ with the Nachtmann variable
$\xi=2x/(1+\sqrt{1+4x^2M_N^2/Q^2})$. This also
shares the above ``unphysical region'' problem. Lastly, working in
Collinear Factorization (CF) the kinematic boundaries are naturally
respected. One advantage of the CF formalism
versus the OPE formalism is that the former can be also applied to
semi-inclusive DIS \cite{Accardi:2009md}, and indeed to any
hard-scattering process. An application to 
parity-violating DIS was
discussed by T.~Hobbs \cite{Hobbs-HiX2010}. 

However, TMCs do not exhaust all possible $O(1/Q^2)$ power
corrections. These include dynamical higher-twist corrections (parton
correlations) as 
well as all uncontrolled leading-twist power corrections, such as Jet Mass
Corrections \cite{Accardi:2008ne}. They also include higher-order perturbative terms, which
are logarithmic in $Q^2$ but  resemble a power law at low 
scales, and large-$x$ resummation effects
\cite{Liuti-HiX2010}. Despite their disparate origin, it is 
customary to label these ``residual'' corrections as ``higher-twist'',
as I will do here.
In the CTEQ6X fits, these HT corrections are parameterized
phenomenologically using a multiplicative factor modifying the
structure function of the proton and the neutron:   
\begin{align}
  F_2^{\,data}=F_2^{TMC}(1+C(x)/Q^2) \ ,
\end{align}
where $F_2^{TMC}$ denotes the structure function after the target mass
corrections have been made. The function $C(x)$ is given by
$
  C(x) = a x^b (1 + cx)
$.
After inclusion of TMCs, this parameterization is sufficiently flexible
to give a good description of the data. To simplify the global fits,
the HT corrections for protons and neutrons where taken equal, given
that their difference was found to be relatively small in other studies.

\section{Nuclear corrections}

In order to separate the $d$ and $u$ quark at large $x\gtrsim0.6$ it
is necessary to consider DIS data on deuteron targets, which are
sensitive to a different linear combination of $u$ and $d$ quarks than
the corresponding data on proton targets. However, at large $x$ the
deuteron deviates from a simple sum of a free proton and neutron due
to significant effects of nuclear binding, Fermi motion and nucleon
off-shellness \cite{Kulagin-HiX2010}. 

\begin{figure}
  \includegraphics[scale=0.35,bb=18 184 592 718,clip=true]
                  {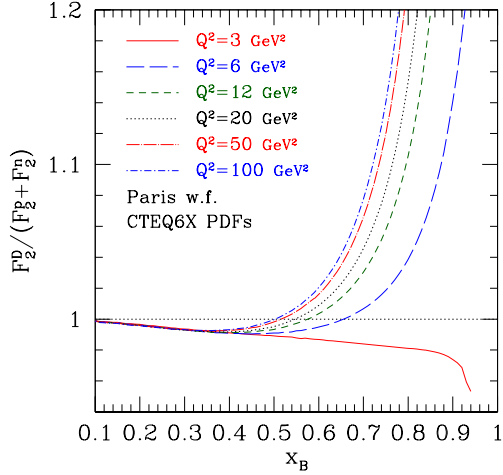}
  \caption{
    $Q^2$ dependence of the Deuteron correction factor 
    $F_2^D/(F_2^p+F_2^n)$ calculated with CTEQ6X PDFs, TMCs in the CF 
    prescription, the fitted HT corrections, and the Paris deuteron wave
    function. 
}
\label{fig:Dpn}
\end{figure}

Since the deuteron is weakly bound one can approximate the bound
nucleon structure function by its on-shell value, and write
the deuteron ($d$) structure function as 
\begin{eqnarray}
  F_2^d(x,Q^2)
  &\approx& \sum_{N=p,n} \int dy\
    f_{N/d}(y,\gamma)\ F_{2N}^{TMC+HT}\left(\frac{x}{y},Q^2\right)\, .
\label{eq:F2d}
\end{eqnarray}
Here $F_{2N}$ is the nucleon (proton $p$ or neutron $n$) structure
function including TMC and HT corrections. The ``smearing
function'' $f_{N/d}$ is computed from the deuteron wave
function and implements nuclear binding and Fermi motion corrections;
at $Q^2\rightarrow\infty$ it can be interpreted as the light-cone 
momentum distribution of nucleons in the deuteron. 
The variable $y = (M_d/M_N) (p_N \cdot q/ p_d \cdot q)$ is the
deuteron's momentum fraction carried by the struck nucleon,
where $q$ is the virtual photon four-momentum, $p_{N(d)}$ the nucleon
(deuteron) four-momentum,  and $M_d$ the deuteron mass; it differs from the light-cone fractional momentum by terms of
$O(M_N^2/Q^2)$. 
Off-shell 
corrections to $F_{2N}$  can also be implemented in
%this formalism 
under a few assumptions, which are outside the
scope of this talk.

The deuteron correction factor $F_2^d/(F_2^p+F_2^n)$ computed with the
Paris wave function and the CTEQ6X PDFs is plotted in
Figure~\ref{fig:Dpn}. It shows a remarkable $Q^2$ dependence at
$Q^2\lesssim 20$ GeV. Part of this $Q^2$ dependence comes from the
smearing function $f_{N/D}$, which depends on
the target mass through the variable 
$\gamma^2=1+4x^2M_N^2/Q^2$. However, this induces only minor
effects on the deuteron correction factor. 
Most of the $Q^2$ dependence shown in the figure is due to TMC and HT
corrections at the nucleon level.

It is also clear that nuclear smearing corrections do not
disappear at large $Q^2$: in general, they are not a subleading
effect, and cannot be avoided by kinematics cuts such as those
commonly used in global PDF fits.

\section{The CTEQ6X parton distributions at large $\bm x$}

The CTEQ6X global PDF fits \cite{Accardi:2009br} were performed at NLO
to a wide variety of data 
similar to that used in the determination of the CTEQ6M1 
PDFs except that no neutrino data were used since their use would
involve additional nuclear corrections beyond those for deuterium. In
addition, the E-866 dimuon data were added as were data for the CDF 
$\gamma+jet$ production, the CDF $W$ lepton asymmetry, and
the D\O\ $W$ asymmetry. 

\begin{figure}
\includegraphics[scale=0.43,bb=18 324 332 718,clip=true]
                {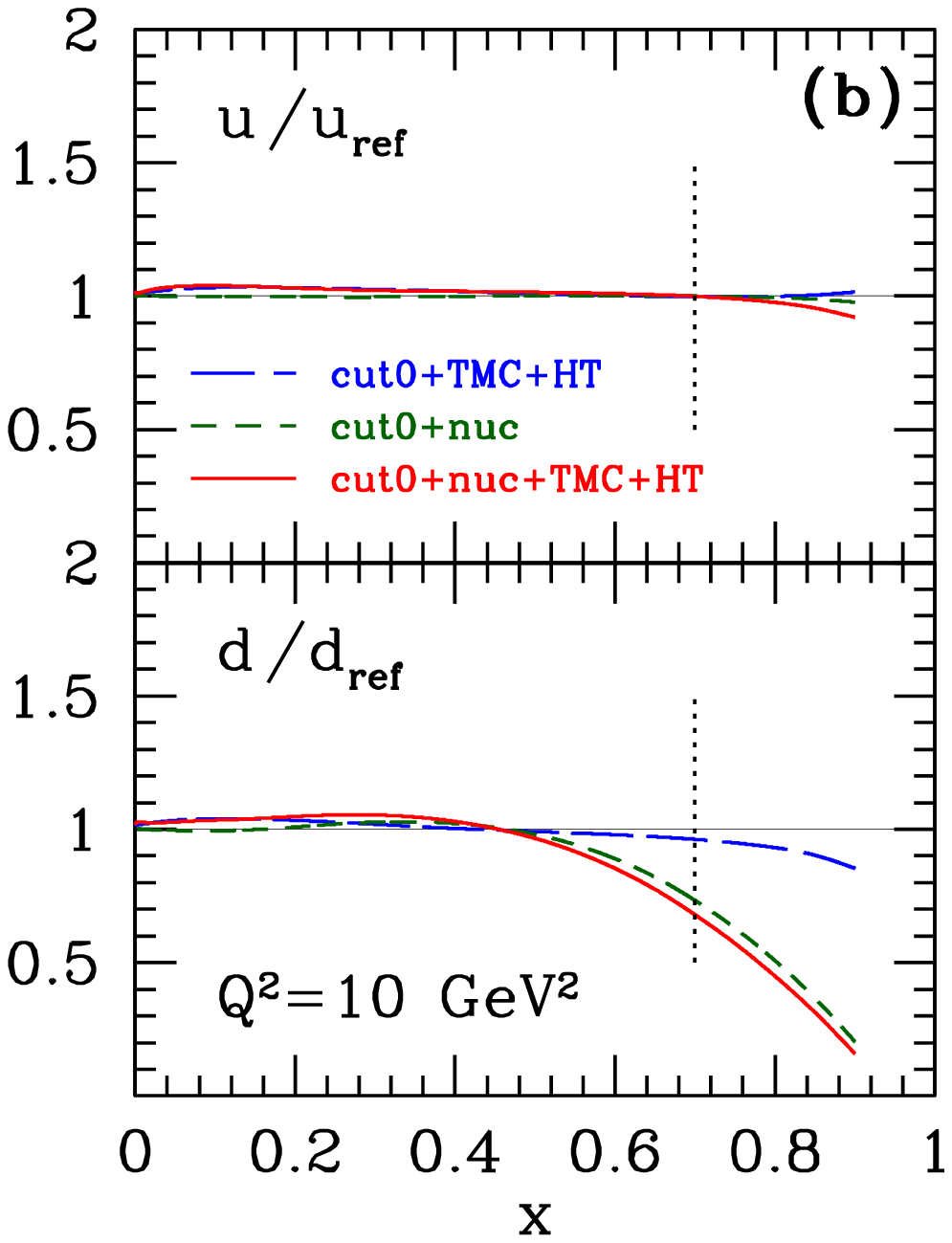}
\includegraphics[scale=0.43,bb=18 364 592 718,clip=true]
                {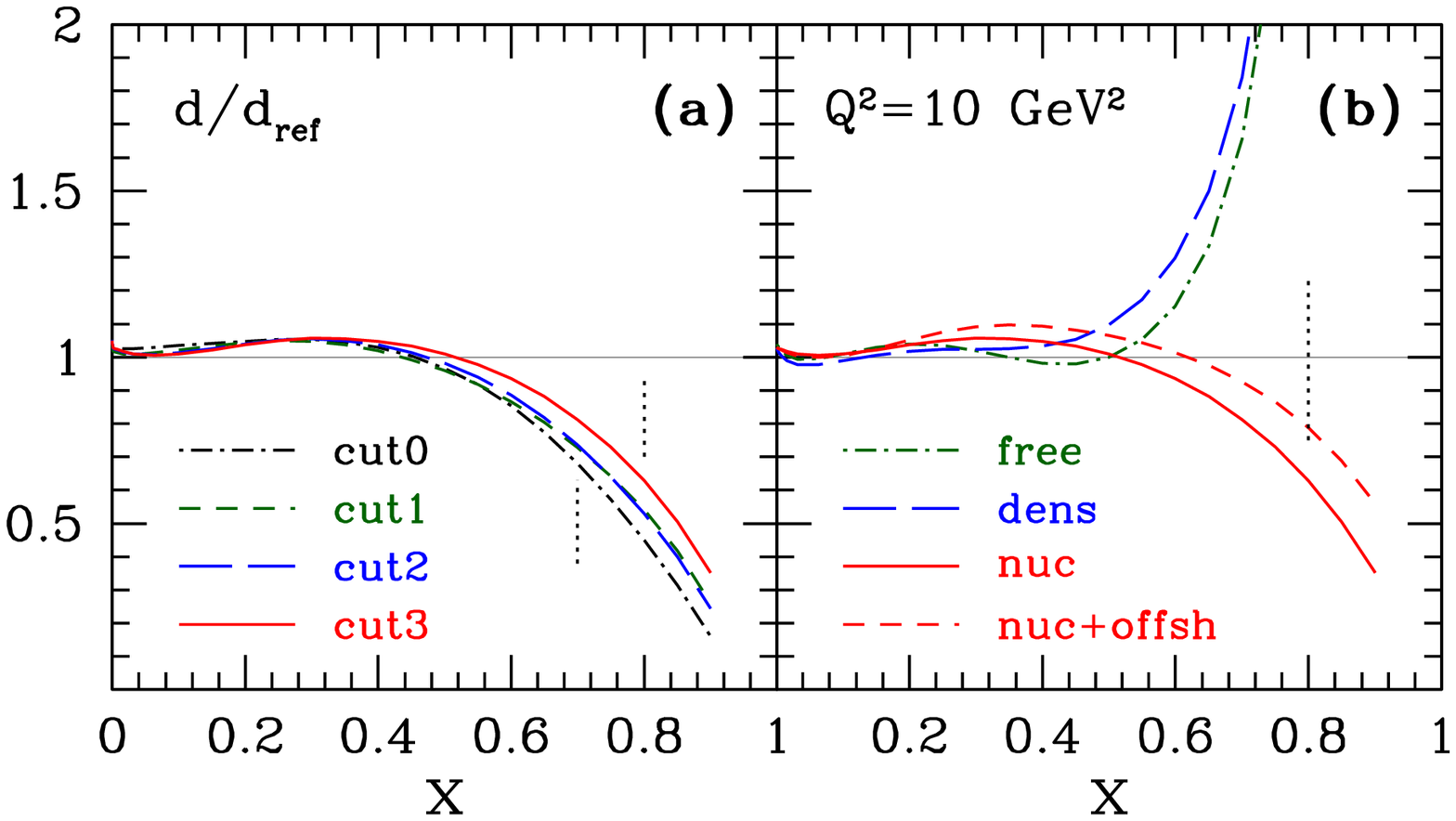}
\caption{
  Left: PDF fits with standard DIS kinematic cuts and different
  combinations of TMC, HT and nuclear corrections. Right: sensitivity
  of the $d$ PDF to kinematic cuts and nuclear corrections. Plots from
  Ref.~\cite{Accardi:2009br}.
}
\label{fig:cut0-cut3}
\end{figure}

Initially, a reference fit (``ref'') was done using
the standard $W > 3.5$ GeV and $Q>2$ GeV (labeled \texttt{cut0}), 
with TMC, HT and nuclear
corrections turned off in order to compare to the CTEQ6M1 PDFs.
The E866 data favor at large $x$ a slightly reduced $u$ PDF and an
increased $d$ PDF; however in the latter case the $W$ asymmetry and
$\gamma$+jet data compensate the increase leaving the $d$ PDF nearly
unchanged. 

Subsequently, several prescriptions for TMC, HT and nuclear
corrections were considered, and the
DIS kinematic cuts progressively relaxed to $W>1.73$ GeV and $Q>1.3$ GeV,
in order to avoid most of the resonance region but be able to include
a good number of Jefferson Lab data. This cut is labeled
\texttt{cut3}, with intermediate cuts labeled \texttt{cut1} and
\texttt{cut2}. 
As TMCs, the discussed OPE, $\xi$-scaling
and CF prescriptions were considered. Nuclear smearing was performed
using the Paris wave function with on-shell nucleon structure
functions or off-shell corrections from the MST model
\cite{Melnitchouk:1994rv};  
the results were compared to fits obtained using
either no nuclear corrections apart from isospin 
effects, or
nuclear corrections in the Density Model, which
extrapolates the nuclear effects observed in heavier nuclei to the
deuteron. 

\begin{figure}
\includegraphics[scale=0.35,bb=18 175 592 700,clip=true]
                {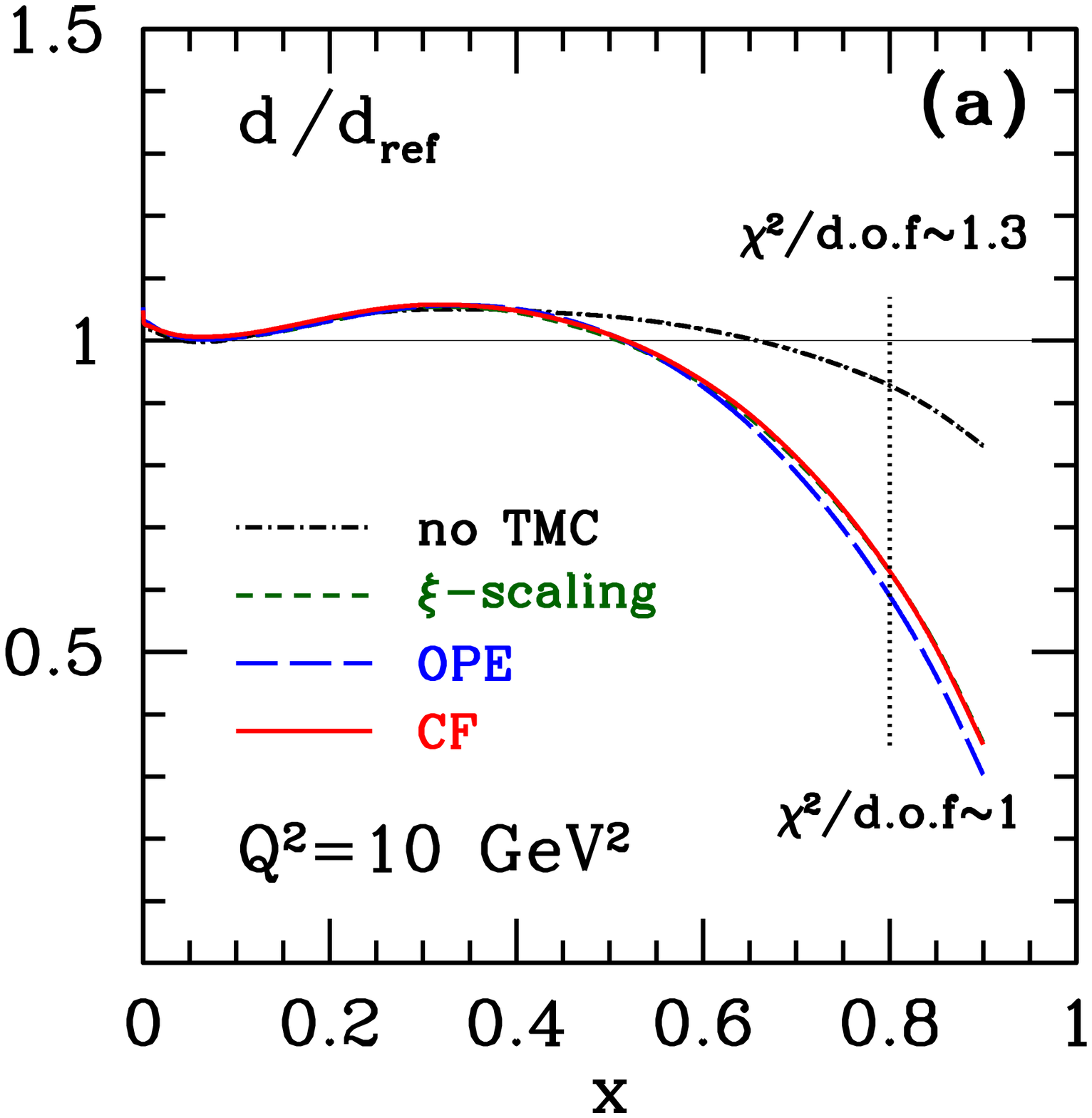}
\includegraphics[scale=0.35,bb=18 175 592 700,clip=true]
                {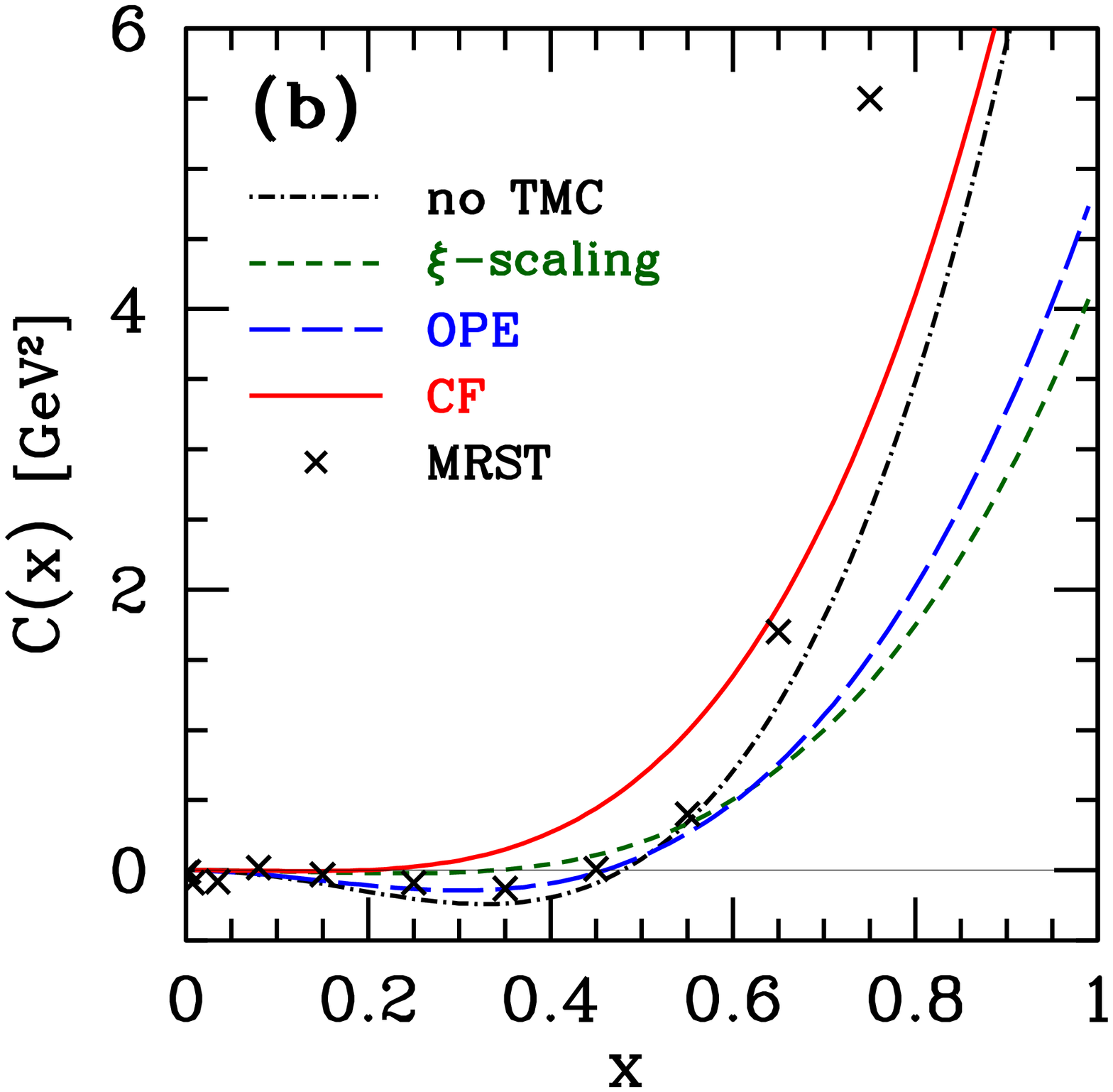}
\caption{
  Sensitivity of the $d$ PDF (left) and fitted HT term
  (right) to various TMC prescriptions.Plots taken from
  Ref.~\cite{Accardi:2009br}. 
}
\label{fig:TMC-HT}
\end{figure}

The main results of the CTEQ6X analysis can be summarized as 
follows.
\begin{itemize}
\item {\bf Standard kinematic cuts.} 
  When using the standard DIS kinematic cuts, $W>3.5$ GeV and
  $Q>2$ GeV, the PDFs are insensitive to TMC and HT corrections;
  however nuclear corrections are large and start at $x\gtrsim 0.45$,
  in a region well inside what is included in the cuts
  (Fig.~\ref{fig:cut0-cut3}, left). 
\item {\bf Enlarged kinematic cuts.} 
  The PDFs are relatively stable against variations of the DIS cuts
  in the vicinity of the $W>1.73$ GeV and $Q>1.3$ cut
  (Fig.~\ref{fig:cut0-cut3}, right plot, left panel). 
  As a consequence of the enlarged data set, there is a
  substantial reduction in the uncertainty of these PDFs due to the
  increased data, with the {\tt cut3} errors reduced by 10--20\% for
  $x \lesssim 0.6$, and by up to 40--60\% at larger $x$.
\item {\bf Stability with respect to TMCs.} 
  The PDFs are nearly independent of the TMC prescription
  (Fig.~\ref{fig:TMC-HT}, left); this is very important for fitting
  leading-twist PDFs. Changes in TMCs are absorbed  by
  the phenomenological HT term, for which TMC modeling induces a
  non-negligible systematic uncertainty 
  (Fig.~\ref{fig:TMC-HT}, right). 
\item {\bf Large sensitivity to nuclear corrections.}
  The $d$ PDF is very sensitive to the nuclear
  correction model adopted (Fig.~\ref{fig:cut0-cut3}, right plot,
  right panel). This induces a large systematic uncertainty, further 
  discussed in the next section.
\end{itemize}

\section{Nuclear uncertainties and collider physics}

A detailed investigation of
the systematic PDF uncertainties induced by nuclear corrections
modeling is underway, as reviewed by J.~Owens
\cite{Owens-HiX2010}. One very
important result is that, surprisingly, the large-$x$ {\em
  gluon} PDF is as 
sensitive to nuclear corrections as the $d$ PDF; they are
anticorrelated to each other due to an interplay of DIS, DY and jet data.
%As discussed,
%nuclear smearing induce a large suppression of large-$x$ $d$ quark. This
%is compensated in DY data by a small and opposite variation of
%large-$x$ u quarks. Finally, the variation in jet data, can be
%compensated by a large variation in the {\em gluon} PDF, which turns
%out to be very sensitive to nuclear corrections.

\begin{figure}
\includegraphics[scale=0.60,bb=18 475 592 718,clip=true]
                {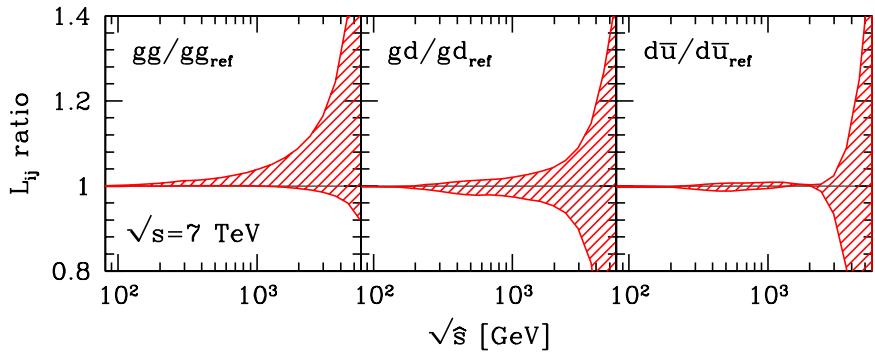}
\caption{
  Relative nuclear uncertainties on parton luminosities at
  $\sqrt{s}=7$~TeV. 
  Shown are the extremes of the variations of the $gg$, $gd$ and
  $d\bar u$ luminosities relative to an intermediate fit (``ref'').
}
\label{fig:luminosities}
\end{figure}

This has potentially profound
implications for future collider experiments, since the resulting
variation of the gluon is significant even at values of $x$ as
low as 0.4. As an illustration, we can consider the parton luminosities,
\begin{align}
  L_{ij} = \frac{1}{(1+\delta_{ij})\hat s} \left[ \int_{\hat{s}/s}^1 \frac{dx}{x}
    f_i\left( x, \hat{s} \right) 
    f_j\left( \frac{\hat{s}}{xs}, \hat{s} \right) 
    + i \leftrightarrow j \right],
\end{align}
where $s$ ($\hat{s}$) is the hadronic (partonic) center of mass
energy squared, and $f_i$ is the PDF for a parton of flavor
$i$ at $Q^2=\hat s$. As an example, 
the $gg$ luminosity controls the total main channel for Higgs production,
the $gd$ luminosity controls the ``standard candle'' cross section
for $W^-$ production, and the $d\bar u$ luminosity is relevant to jet
production.
These are plotted in Fig.~\ref{fig:luminosities}
for $\sqrt{s}=7$~TeV, relevant to the current LHC runs. The nuclear
uncertainty 
%starts at rather small scales $\sqrt{\hat{s}}$, and 
grows quickly above 5-10\% as $\sqrt{\hat{s}}$ exceeds 1~TeV.

\section{An application to quark-hadron duality}

\begin{figure}
  \includegraphics[scale=0.5,bb=0 25 567 550,clip=true]{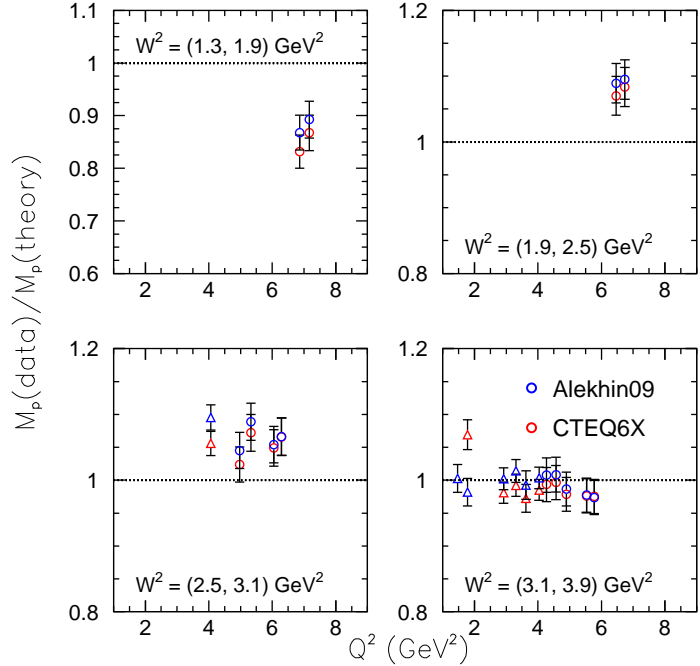}
  \caption{
    Preliminary results of a comparison of averaged JLab $F_2$ data
    over 4 ``resonance regions'' and computations using CTEQ6X (red) and
    Alekhin09 PDFs (blue).
}
\label{fig:QHD}
\end{figure}

Quark-hadron duality in structure functions refers to the experimental
observation that inclusive structure functions in the
region dominated by low-lying nucleon resonances follow deep inelastic
structure functions describing high energy data, to which the resonance
structure functions average \cite{Melnitchouk-HiX2010}. 
The new large-$x$ CTEQ6X PDFs can be used to verify to what degree
this holds true, which is important to understand the
transition from the perturbative (partonic) to the non perturbative
(hadronic) regime of QCD.
  
The handbag diagram used in the pQCD computations assumes no
interaction between the scattered quark and the 
target remnant. This is a reasonable approximation only if their
rapidity separation, $\Delta y$, is large enough. A value of
$\Delta y > 2-4$, known as ``Berger criterion''
\cite{Berger-Mulders}, should be sufficient to ensure applicability of
pQCD. Since
$\Delta y$ decreases as $x \rightarrow 1$ or $Q^2\rightarrow
0$, this limits the range in $x$ and $Q^2$ where
the comparison of pQCD computations and resonance region data makes sense. 

In Figure~\ref{fig:QHD} the ratios of Jefferson Lab data
averaged over different resonance regions to
computations using CTEQ6X are plotted
\cite{Malace-HiX2010,Accardi-Malace-prep}. These are compared to 
calculations using Alekhin09 PDFs \cite{Alekhin-HiX2010}, which were
fitted with similar TMC, HT and nuclear corrections as CTEQ6X.
Only data satisfying a conservative $\Delta y>4$ are
retained. Data in the $W^2=(3.1,3.9)$ GeV$^2$ region were included in the
CTEQ6X fit, and indeed are well described. The $W^2=(2.5,3.1)$ GeV$^2$ 
region, while not directly included in the fit, is constrained by DIS data
at larger $Q^2$ because of DGLAP evolution.
Data in the $W^2=(1.9,2.5)$ GeV$^2$ region
lie in the extrapolation region of the CTEQ6X fits.

The plot shows that quark-hadron duality works within 5-10\%. This opens
the possibility of using resonance region data to extend the range of
validity in $x$ of the fits. A study is underway
\cite{Accardi-Malace-prep} to further explore 
this issue, e.g., by determining whether duality holds for smaller
values of the $\Delta y$ cut, which would enlarge the number of data
points, by evaluating nuclear systematic uncertainties and by quantifying
PDF uncertainties from the HT parameters, especially in the
extrapolation region.

\section{Conclusions}

I have shown that a good control of global PDF fits can be achieved
when including DIS data in the pre-asymptotic region of large $x$
and small $Q^2$, if one considers TMC and HT corrections. 
The resulting PDFs are stable against available TMC prescriptions,
with the modeling uncertainty absorbed in the phenomenologically
extracted HT terms. This is very good for applications to collider and
neutrino physics \cite{Morfin-HiX2010}, and for comparing PDF moments to lattice
calculations \cite{Renner-HiX2010,Detmold-HiX2010}.

Theoretical nuclear corrections to deuteron target data are
necessary for $u$ and $d$ quark separation at $x\gtrsim0.5$.
The $d$-quark and, surprisingly, the
gluon PDF turn out to be very sensitive to uncertainties in the
modeling and calculation of nuclear effects. 
Furthermore, the induced uncertainty on parton
luminosities at the LHC is non-negligible.

A careful study shows that the induced systematic 
$d$-quark PDF uncertainty is of the same order as the experimental
uncertainty if the deuteron target data are removed from the
fit \cite{Owens-HiX2010}. 
Therefore, further progress in constraining the $d$ quark and the
gluon PDFs at large
$x$ requires a better theoretical understanding of nuclear
corrections, combined with new data on free proton, but sensitive to
$d$, such as from parity violating DIS or from neutrino DIS on a
hydrogen target \cite{Morfin-HiX2010}. Alternatively one can
use data minimizing
nuclear corrections, such as from proton-tagged DIS on deuteron targets, for
which the nuclear uncertainty is smaller than 2\%
\cite{Accardi-Melnitchouk-prep}. 
%See \cite{Owens-HiX2010,Accardi:2009br} for more details. 
Using quark-hadron duality to include 
resonance region data in the fits, thereby extending their $x$ range,
seems also feasible.
Finally, data sensitive to large $x$ gluons, such as from the
longitudinal and charm structure functions, $F_L$ and $F_2^c$, are
required to constrain the gluons independently of the jet data.

%%%%%%%%%%%%%%%%%%%%%%%%%%%%%%%%%%%%%%%%%%%%%%%%
%% BACKMATTER
%%%%%%%%%%%%%%%%%%%%%%%%%%%%%%%%%%%%%%%%%%%%%%%%

\begin{theacknowledgments}

I am most grateful to 
	M.~E.~Christy
	C.~E.~Keppel,
        S.~Malace,
	W.~Melnitchouk,
	P.~Monaghan,
	J.~G.~Morf\'\i n,	
	J.~F.~Owens, and
        L.~Zhu
for their collaboration on the matter presented in this talk.
This work has been supported by the DOE contract DE-AC05-06OR23177,
under which Jefferson Science Associates, LLC operates Jefferson Lab,
and NSF award No.~1002644.
\end{theacknowledgments}

%%%%%%%%%%%%%%%%%%%%%%%%%%%%%%%%%%%%%%%%%%%
%% The following lines show an example how to produce a bibliography
%% without the help of the BibTeX program. This could be used instead
%% of the above.
%%%%%%%%%%%%%%%%%%%%%%%%%%%%%%%%%%%%%%%%%%%


\begin{thebibliography}{99}

\bibitem[Owens (2010)]{Owens-HiX2010}
  J.~Owens, A.~Accardi, C.~E.~Keppel, M.~E.~Christy, W.~Melnitchouk,
  P.~Monaghan,, L.~Zhu, J.~G.~Morf\'\i n, 
  ``Global Fits for PDFs'', these proceedings.

\bibitem[Guffanti (2010)]{Guffanti-HiX2010}
  A.~Guffanti, ``PDFs and neural networks'', these proceedings.

\bibitem[Radescu (2010)]{Radescu-HiX2010}
  V.~Radescu, ``Parton distributions from HERA'', these proceedings.

\bibitem[Holt (2010)]{Holt-HiX2010}
  R.~Holt, ``Overview of structure function measurements at large $x$'',
  these proceedings. 

\bibitem[Steffens (2010)]{Steffens-HiX2010}
  F.~Steffens, ``New approaches to target mass corrections'', these
  proceedings. 
 
\bibitem[Liuti (2010)]{Liuti-HiX2010}
  S.~Liuti, ``$W$ evolution at large $x$'', these proceedings.

\bibitem[Glatzmaier (2010)]{Glatzmaier-HiX2010}
  M.~Glatzmaier, ``Higher twist scaling violations'', these
  proceedings.

\bibitem[Lee (2010)]{Lee-HiX2010}
  S.-H.~Lee, ``Higher twists in DIS'', these proceedings.
 
\bibitem[Melnitchouk (2010)]{Melnitchouk-HiX2010}
  W.~Melnitchouk, ``Quark-hadron duality in structure functions'',
  these proceedings. 

\bibitem[Malace (2010)]{Malace-HiX2010}
  S.~P.~Malace, ``Neutron and proton structure functions and
  duality'', these proceedings. 

\bibitem[Accardi et al. (2010)]{Accardi:2009br}
  A.~Accardi, M.~E.~Christy, C.~E.~Keppel, P.~Monaghan, W.~Melnitchouk, J.~G.~Morf\'\i n and J.~F.~Owens,
  %``New parton distributions from large-x and low-Q^2 data,''
  Phys.\ Rev.\  D {\bf 81}, 034016 (2010).
  %[arXiv:0911.2254 [hep-ph]].
  %%CITATION = PHRVA,D81,034016;%%

\bibitem[Accardi et al. (2009)]{Accardi:2009md}
  A.~Accardi, T.~Hobbs and W.~Melnitchouk,
  %``Hadron mass corrections in semi-inclusive deep inelastic scattering,''
  JHEP {\bf 0911}, 084 (2009).
  %[arXiv:0907.2395 [hep-ph]].
  %%CITATION = JHEPA,0911,084;%%

\bibitem[Hobbs (2010)]{Hobbs-HiX2010}
  T.~Hobbs, ``Finite-Q2 corrections in PVDIS'', these proceedings.

\bibitem[Accardi, Qiu (2008)]{Accardi:2008ne}
  A.~Accardi and J.~W.~Qiu,
  %``Collinear factorization for deep inelastic scattering structure functions
  %at large Bjorken xB,''
  JHEP {\bf 0807}, 090 (2008).
  %[arXiv:0805.1496 [hep-ph]].
  %%CITATION = JHEPA,0807,090;%%

\bibitem[Kulagin (2010)]{Kulagin-HiX2010}
  S.~Kulagin, ``Nuclear effects in DIS'', these proceedings.

\bibitem[Melnitchouk et al. (1994)]{Melnitchouk:1994rv}
  W.~Melnitchouk, A.~W.~Schreiber and A.~W.~Thomas,
  %``Relativistic deuteron structure function,''
  Phys.\ Lett.\  B {\bf 335}, 11 (1994).
  %[arXiv:nucl-th/9407007].
  %%CITATION = PHLTA,B335,11;%%

\bibitem[Accardi  Malace (2011)]{Accardi-Malace-prep}
S~.P.~Malace and A.~Accardi, {\it in preparation}.

\bibitem[Berger (1979)]{Berger-Mulders}
  %\bibitem{Berger:1987zu}
  E.~L.~Berger, 
  %``Semiinclusive Inelastic Electron Scattering From Nuclei,''
  ANL-HEP-CP-87-45; 
  %%CITATION = C87/01/05.2;%%
  %
  %\bibitem{Mulders:2000jt}
  P.~J.~Mulders,
  %``Current fragmentation in semiinclusive leptoproduction,''
  AIP Conf.\ Proc.\  {\bf 588}, 75 (2001).
  %[arXiv:hep-ph/0010199].
  %%CITATION = APCPC,588,75;%%

\bibitem[Alekhin (2010)]{Alekhin-HiX2010}
  S.~Alekhin, ``NNLO PDFs at large x'', these proceedings.

\bibitem[Morfin (2010)]{Morfin-HiX2010}
  J.~Morf\'\i n, ``Looking at High-$x_{Bj}$ with neutrinos'', these proceedings.

\bibitem[Renner et~al. (2010)]{Renner-HiX2010}
  D.~Renner, ``PDF moments in lattice QCD'', these proceedings.

\bibitem[Renner et~al. (2010)]{Detmold-HiX2010}
  W.~Detmold, ``Higher moments of PDFs in lattice QCD'', these proceedings.

\bibitem[Accardi, Melnitchouk (2011)]{Accardi-Melnitchouk-prep}
  A.~Accardi, W.~Melnitchouk, {\it in preparation}.

\end{thebibliography}
\end{document}